\documentclass[12pt]{iopart}
\usepackage{iopams}
\usepackage{epsfig}
\begin{document}
\jl{1}
\newcommand{\n}{{\bi n}}
\newcommand{\nbar}{\bar{\bi n}}
\newcommand{\Linf}{L\rightarrow\infty}
\title[Wigner function from path integral]{Coherent-state path integral calculation of the Wigner function}
\author{J H Samson\footnote[1]{Electronic address: 
j.h.samson@lboro.ac.uk}}
\address{Department of Physics, Loughborough University, 
Loughborough, Leics LE11 3TU, United Kingdom}
\begin{abstract}
We consider a set of operators $\hat{\bi x}=(\hat{x}_1,\ldots,
\hat{x}_N)$ with diagonal representatives ${\bi P}({\bi n})$ in the
space of generalized coherent states $|\n\rangle:\; \hat{\bi x}=\int
\rmd \mu(\n) {\bi P}({\bi n}) |\n\rangle \langle\n|$.  We regularize
the coherent-state path integral as a limit of a sequence of averages
$\langle\;\rangle_{L}$ over polygonal paths with $L$ vertices ${\bi
n}_{1\ldots L}$.  The distribution of the path centroid $\bar{\bi
P}=\frac{1}{L}\sum_{l=1}^{L}{\bi P}({\bi n}_{l})$ tends to the Wigner
function $W({\bi x})$, the joint distribution for the operators:
$W({\bi x})=\lim_{L\rightarrow\infty}\langle\delta_{N}({\bi
x}-\bar{\bi P})\rangle_{L}$.  This result is proved in the case where
the Hamiltonian commutes with $\hat{\bi x}$.  The Wigner function is
non-positive if the dominant paths with path centroid in a certain
region have Berry phases close to odd multiples of $\pi$.  For finite
$L$ the path centroid distribution is a Wigner function convolved with
a Gaussian of variance inversely proportional to $L$.  The results are
illustrated by numerical calculations of the spin Wigner function from
SU(2) coherent states.  The relevance to the quantum Monte Carlo sign
problem is also discussed.
\end{abstract}

\submitted Version 28 March 2000.  Accepted 5 June 2000.
%\maketitle

\section{Introduction} \label{intro}
\begin{quote}
	\emph{It is possible, therefore, that a closer study of the 
	relation of classical and quantum theory might involve us in 
	negative probabilities, and so it does.} R~P~Feynman\cite{Feynman}
\end{quote}

The application of classical concepts to quantum systems comes at a
cost.  It is indeed possible to represent operators in many ways by
functions $f$ of commuting variables. 
Expectation values take a form reminiscent of classical
statistical mechanics: they are averages with respect to a
distribution $W({\bi x})$, where $\bi x$ takes values in some (as yet 
unspecified) space $\Gamma$:
\begin{equation}
	\langle \hat{f} \rangle = \Tr (\hat{f}\hat{\rho}) = 
	\int_{\Gamma} 
	\rmd {\bi x} f({\bi x}) W({\bi x}).
	\label{eq:expval}
\end{equation}
Here $f$ is a function that depends only on the operator $\hat{f}$ of
interest, and $W$ is a normalized distribution that depends only on
the density matrix $\hat{\rho}$ \cite{Strat}.  Although \emph{quasiprobability} distributions
$W({\bi x})$ satisfying \eref{eq:expval} indeed exist \cite{HOSW,Mueckenheim}, it
is often the case that no \emph{positive-definite} $W({\bi x})$ exists
for a given $\hat{\rho}$.  The
distribution that is the principal subject of this paper is the Wigner
function, best known as a joint distribution of position and
momentum \cite{HOSW}. Another example is the distribution of local hidden
variables that predetermine the outcome of measurements on components 
of an entangled state.  Here the assumption $W({\bi
x}) \ge 0$ implies certain inequalities between correlations that are 
violated by the quantum mechanical result, demonstrating the 
nonexistence of a positive distribution \cite{Bell}.

\Eref{eq:expval} also represents path integral methods, where $\bi x$
is a path in a coherent-state manifold or a time-dependent auxiliary
field.  There is no difficulty in principle in working with
non-positive distributions, provided one is sufficiently wary when
applying the axioms of probability theory.  In practice, however,
non-positivity can be a serious hindrance to numerical computation of
the integrals.  Monte Carlo techniques typically evaluate integrals of
the form \eref{eq:expval} by sampling $f(\bi{x})$ from a distribution
$W$.  If the average sign $\int_{\Gamma} W({\bi x})\,\rmd{\bi x} /
\int_{\Gamma} |W({\bi x})|\,\rmd{\bi x} $ is small, convergence with
sample size becomes intolerably slow.  This notorious \emph{sign
problem}, a frequent hindrance to quantum Monte Carlo calculations,
was first noted in the context of fermion simulations \cite{BSS} but
appears in other contexts \cite{review}.  The coherent-state path
integral presents the system perhaps least amenable to Monte Carlo
simulation: the weight is complex and contains a rapidly varying Berry
phase.  For this reason, such integrals have rarely \cite{Takano,VS2}
been tackled by Monte Carlo techniques.  A more widely used approach
for Monte Carlo simulation of many-body systems replaces the
interaction by a Gaussian average over an auxiliary field.  The
auxiliary field transports the system around a path in state space,
with the state relaxing towards the instantaneous ground state in the
field \cite{review,Fahy}.  The distribution of the auxiliary field is
a thermally broadened distribution of the operators to which it is 
coupled \cite{S95}.  Thus the results presented here for
coherent states can cast light on the sign problem in the
auxiliary-field Monte Carlo method.

An alternative to direct evaluation of the path integral is to
integrate all non-zero frequency modes out of the path integral,
thereby mapping the system to an effective \emph{classical} system, 
usually determined variationally
\cite{FH,GT,FK}.  Paths ${\bi x}(\tau), 0 \le \tau < \beta$ are
classified according to their \emph{path centroid} ${\bar{\bi x}} =
\frac{1}{\beta} \int_{0}^{\beta}{\bi x}(\tau){\mathrm d}\tau$.  The
path integral with action ${\cal S}[\bi{x}]$ reduces to an ordinary
integral over $c$-number variables with classical effective
Hamiltonian $H_{\mathrm{eff}}(\bar{\bi x})$.  Excursions of the path
from $\bar{\bi x}$ provide quantum corrections to the potential in the
effective Hamiltonian.  The function $W(\bar{\bi x})=\exp(-\beta
H_{\mathrm{eff}}(\bar{\bi x}))$, regarded as a Boltzmann
distribution of the variables $\bar{\bi x}$, forms the basis of
a classical statistical mechanics in phase space \cite{CV,CGTVV}.  This distribution is an intuitive
interpretation of the path centroid distribution \emph{if the
observables represented by $\bi{x}$ are compatible}.  However, this is
not always the case: the distribution resulting from a phase-space
path integral over canonical coordinates $(q,p)$ would be a Wigner
function.  It is such distributions that form the subject of 
this paper.

The present author has previously shown a formal correspondence
between the path-centroid spin distribution in the coherent-state path
integral and the spin Wigner function \cite{S95}.  This work made the
assumption of continuous paths, implicit in field-theoretical
treatments \cite{FS}, leading to representation of the spin by its
matrix element ${\bi Q}=s\n$.  (Here the coherent-state label $\n$ is
a unit vector in ${\mathbb R}^{3}$.)  However, because of the
conditionally convergent nature of the path integral, the
correspondence between the spin representation in the path integral
and the Wigner function depends on the class of paths appearing in the
path integral.  While some forms of the measure require use of the
matrix element \cite{Klauder}, Brownian paths in the limit of
divergent diffusion coefficient are non-differentiable, and require
the diagonal representative ${\bi P}=(s+1)\n$ \cite{DK}.  The apparent
ambiguity of the functional integral representations of the Hubbard
model has similarly been ascribed to questions of continuity of paths
\cite{Prange}.  Subsequent work reported briefly \cite{S99}
demonstrated that, if the path integral for spin is defined on a
sequence of spherical polygons, the distribution of the path centroid of the diagonal
representative of the spin converges onto the spin Wigner function. 
The aim of the present work is to generalize this result to
coherent-state representations in arbitrary finite-dimensional Hilbert
spaces, and discuss the convergence with
path discretization.

The next section defines the Wigner function and coherent state
formalism used here.  \Sref{sec:calc} derives the main result of this
work, that the Wigner function of a set of operators can be computed
as a histogram of their time-averaged diagonal representatives:
\begin{equation}
	W({\bi x})\equiv\langle\delta_{N}({\bi x}-\hat{\bi x})\rangle=
	\lim_{L\rightarrow\infty}\langle\delta_{N}({\bi
	x}-\bar{\bi P})\rangle_{L}.
	\label{eq:result}
\end{equation}
Here $\hat{\bi x}$ are operators commuting with the Hamiltonian, $\bi x$ are $c$-number variables and
$\bi P$ are the diagonal representatives of the operators $\hat{\bi
x}=\int \rmd \mu(\n) {\bi P}({\bi n}) |\n\rangle \langle\n|$.  $\bar\bi P$
is the path centroid, the time average of the diagonal representative.  The first set of angle brackets represents a
thermal expectation value, and the second an average over $L$-vertex 
polygonal paths in the
coherent-state path integral; $\delta_{N}$ is the delta function in 
${\mathbb R}^{N}$.  For finite $L$ the distribution is broadened,
typically by a Gaussian of variance inversely proportional to $L$.  
An application to a spin $s$ particle is presented in \sref{sec:spin}.  
Finally, \sref{sec:conc} discusses the interpretation and wider
applicability of these results.  Non-positive regions of the Wigner 
function tend to emerge if the dominant path for a given 
value of the path centroid has a Berry phase that is an odd multiple 
of $\pi$.

\section{Definitions}
\label{sec:def}
\subsection{The Wigner function}
The state of a system is specified by its density matrix
$\hat{\rho}=Z^{-1}\exp(-\beta \hat{H})$, a positive Hermitian operator
with trace $\Tr\hat{\rho}=1$.  This is written in the form of a
canonical density matrix of a system in thermal equilibrium at an
inverse temperature $\beta$; the foregoing equation can be taken as
defining a Hamiltonian (up to an additive constant) for any
non-singular $\hat{\rho}$.  We map the system onto the statistical
mechanics of $c$-number variables ${\bi
x}\equiv(x_{1},\ldots,x_{N})\in{\mathbb {R}}^{N}$ corresponding to $N$
linearly independent operators $\hat{\bi
x}\equiv(\hat{x}_{1},\ldots,\hat{x}_{N})$.  The quasiprobability
distribution $W(\bi{x})$ is the \emph{Wigner function} \cite{HOSW} of
the corresponding operators.  This function could be considered as a
Boltzmann distribution with a (not necessarily real or bounded)
classical effective Hamiltonian $H_{\mathrm{eff}}(\bi{x})$.  We
require $W(\bi{x})$ to be a linear function of $\hat{\rho}$, defined
as
 \begin{equation}
  	W({\bi x}) = \Tr \left(\hat{\rho}\, 
  	  	  	\delta_{N}({\bi x}-\hat{\bi x})\right) ,
  	\label{eq:Weyl}
  \end{equation}
  where the $N$-dimensional delta function $\delta_{N}$ is defined
  with symmetrical operator ordering \cite{SW}
  \begin{equation}
		\delta_{N}({\bi x}-{\hat{\bi x}}) \equiv 
		\int\frac{{\rm d}^{N}\blambda}{(2\pi)^{N}}
		\exp\left(\rmi\blambda\cdot({\bi x}-{\hat{\bi x}})\right).
    	\label{eq:delta}
    \end{equation}
The Wigner function is then the Fourier transform of the 
\emph{characteristic function}
\begin{eqnarray}
				\chi(\blambda)& = & \Tr \left( 
				\hat{\rho}\,\rme^{-\rmi\blambda\cdot\hat{\bi 
				x}}\right): \label{eq:chi} \\
	W({\bi x}) & = & \int\frac{{\rm d}^{N}\blambda}{(2\pi)^{N}}
		\rme^{\rmi\blambda\cdot{\bi x}}\chi(\blambda).
	\label{eq:W}
\end{eqnarray}
It can easily be shown that the Wigner function \eref{eq:W} reduces to
the correct positive marginal distribution in an $M$-dimensional
commuting subspace of operators on integration over the remaining
$N-M$ variables.  The Wigner function can therefore be used to find
expectation values of linear combinations of arbitrary functions of
commuting operators.  The correlation between two spins, $\langle
\hat{\bi S}_{1}\cdot {\hat{\bi S}}_{2}\rangle = \sum_{i=x,y,z}\langle
\hat{S}^{(i)}_1\hat{S}^{(i)}_2 \rangle $, is of this form, as are the
correlations $\langle{\bi a} \cdot \hat{\bi S}_{1}{\bi b} \cdot
\hat{\bi S}_{2}\rangle$ measured in the Einstein-Podolsky-Rosen
experiment \cite{Bell,Agarwal}.

If ${\hat{\bi x}}=(\hat{q},\hat{p})$ are the canonical position and
momentum operators, \eref{eq:W} reduces to the well-known form of the
Wigner function.  If ${\hat{\bi x}}=\hat{\bi S}$ is the 
spin operator for a spin-$s$ particle, we obtain a rather singular 
form consisting of derivatives of 
$\delta$-functions supported on spheres of quantized radius.  In zero 
field this is \cite{S95,Chandler,SW}
\begin{equation} 
	W_{s}({\bi S}) = \left\{
		\begin{array}{cc}
				\frac{-1}{2s+1}\sum_{m=1/2}^{s}\frac{1}{2\pi 
			S}\delta^{\prime}(S-m),	 & s \mathrm{~half~odd~integer}  \\ & \\
			\frac{\delta_{3}({\bi S})}{2s+1} - 
			\frac{1}{2s+1}\sum_{m=1}^{s}\frac{1}{2\pi 
			S}\delta^{\prime}(S-m), & s \mathrm{~integer}
				\end{array}
			\right. .
	\label{eq:evenodd}
\end{equation}
(Here $S=|{\bi S}|$ is the magnitude of the classical spin vector, and $s$ is the spin 
quantum number.)

The representation of the spin distribution as a function of a vector
${\bi S} \in {\mathbb R}^{3}$ demands comment.  This is appropriate in
the context of the statistical mechanics of composite spins, where the
state is not restricted to a single spin-$s$ representation.  The total spin has a distribution
in the same space, which allows for dispersion in the magnitude, and
the formalism is explicitly isotropic.  A more natural choice for a
fixed spin $s$ is the distribution of directions of a vector of
magnitude $\sqrt {s(s+1)}$; the space is a sphere $S^{2}$, and a
family of distributions on the sphere can be derived from the
coherent-state representation of the spin \cite{SW,Agarwal}.  We shall relate the distributions by embedding
the sphere in ${\mathbb R}^{3}$: the correspondence is between sets of
points on $S^{2}$ and their vector average.

\subsection{Coherent states}\label{sec:cs}
Before relating these Wigner functions to coherent-state path
integrals, we review the  properties of generalized coherent states relevant to 
the present work
\cite{KS,Perelomov,ZFG}.
 
Coherent states $|\n\rangle$, labelled by a variable taking values in
some manifold, $\n \in \cal M$, provide a continuous, overcomplete and
non-orthogonal basis for a finite-dimensional Hilbert space $\cal H$.  The completeness 
is expressed by the resolution of the identity
\begin{equation}
	 \hat{1}=\int_{\cal M}\rmd\mu({\n}) 
	|{\n}\rangle \langle {\n}|,
	\label{eq:resolution}
\end{equation}
where $\rmd\mu({\n})$ is a measure on the manifold.
There exists a family of representations of operators by functions on 
the manifold, of which the
matrix element and diagonal representative are of relevance here
\cite{HOSW,BM}.  The \emph{matrix element} $Q$ (also known as the
anti-normal or upper symbol) of an operator $\hat{A}$ is
\begin{equation}
	Q({\n}) = \langle{\n}|\hat{A}|{\n}\rangle,
	\label{eq:Q}
\end{equation}
and the \emph{diagonal representative} $P$ (also known as 
the normal or lower symbol) obeys
\begin{equation}
	\hat{A} = \int_{\cal M} {\rmd}\mu({\n}) P({\n}) |{\n}\rangle \langle 
	\n| .
	\label{eq:P}
\end{equation}
This does not uniquely define the diagonal representative, we only
require that it exist, be bounded and be a linear function of the
operators.  This holds in common cases for suitable choices of the
fiducial vector \cite{MACS}, and can be constructed for SU(2) coherent
states \cite{VS,Gilmore}.

For use in \sref{sec:spin}, we recall the form of 
SU(2) coherent 
states for spin-$s$ as defined by Radcliffe
\cite{Radcliffe}.
Hilbert space ${\cal H}={\mathbb C}^{2s+1}$ is $2s+1$-dimensional and
the coherent-state manifold is the Bloch sphere, ${\cal M}=S^{2}$,
with ${\n} = (\sin\theta \cos\phi, \sin\theta \sin\phi, \cos\theta)$ a
unit vector.  The states are obtained by rotating the highest-weight
eigenstate $|s\rangle$ about an axis in the $xy$ plane:
\begin{equation}
	|\n\rangle  =  \cos^{2s}\frac \theta 2 \exp \left[ 
	\tan\frac \theta 2 \rme^{\rmi\phi}\hat{S}_{-}\right] |s\rangle
	\label{eq:cs}
\end{equation}
This has the useful property that the spin is ``pointing'' in the
direction $\n$, i.~e., $\n \cdot \hat{\bi S}|{\n}\rangle =
s|\n\rangle$.  Such states are not orthogonal:
\begin{equation}
	\langle{\n}_{1}|{\bi n}_{2}\rangle  =  \left( 
	\frac{1+{\n}_{1} \cdot {\bi n}_{2}}{2}\right)^{s}\rme^{is\Omega}, 
	\label{eq:overlap}
\end{equation}
where (with the present gauge) $\Omega$ is the area of the spherical triangle formed by the 
$z$-axis, ${\n}_{1}$ and ${\n}_{2}$.  The measure is the element of 
area ${\rmd}\mu({\bi n}) = \frac{2s+1}{4\pi} {\rmd} 
\cos\theta {\rmd}\phi$.  The matrix element and diagonal 
representatives of the spin operator are ${\bi Q}=s\n$ and ${\bi 
P}=(s+1)\n$ respectively \cite{Lieb}.

\section{Path integral calculation of Wigner function}
\label{sec:calc}
\subsection{Path centroid distributions}
\label{subsec:PCD}

We now derive the main result, relating the Wigner function of an 
operator to the distribution of its diagonal representative, in the 
case of conserved
operators $\hat \bi x$, i.~e.,
\begin{equation}
	[{\hat H},{\hat x}_{\mu}] = 0,\; \mu=1\ldots N.
	\label{eq:cons}
\end{equation}
Thus we can compute the joint distribution of the symmetry generators
of the Hamiltonian, such as the total spin of an isotropic
ferromagnet.  We also require $\hat H$ and $\hat \bi x$ to be bounded
operators, and their diagonal representatives to be bounded functions.

We apply the 
Suzuki-Trotter decomposition to the characteristic 
function \eref{eq:chi},
\begin{equation}
	\chi(\blambda) = \lim_{\Linf}
	Z^{-1}\Tr\left(1-\frac{\beta\hat{H}+\rmi\blambda\cdot\hat{\bi x}}{L}\right)^{L}.
	\label{eq:ST1}
\end{equation}
Let $P_{H}$ be the diagonal representative of the Hamiltonian and
$P_{\mu\nu\cdots}$ be the diagonal representative of the operator
product
\begin{equation}
	\hat{x}_{\mu}\hat{x}_{\nu}\cdots = \int_{\cal M} {\rmd}\mu({\n}) 
	P_{\mu\nu\cdots}({\n}) |{\n}\rangle \langle 
	\n| ,
	\label{eq:Px}
\end{equation}
so that ${\bi P}(\n)=(P_{1}(\n),\ldots,P_{N}(\n))$ is the diagonal 
representative of $\hat\bi x$.
The matrix element clearly exists and is unique for a 
bounded operator.  Replacing the 
operators in each of the $L$ factors in \eref{eq:ST1} by their diagonal representatives gives
\begin{eqnarray}
\fl	\chi(\blambda) & = & \lim_{\Linf} Z^{-1} \Tr \int_{{\cal M}^{L}}\rmd 
\mu_{L}[\n] \prod_{l=1}^{L}\left[\left(1-\frac{\beta 
P_{H}(\n_{l})+\rmi\blambda\cdot {\bi P}(\n_{l})}{L}\right)|\n_{l}\rangle 
\langle\n_{l}|\right]
	\label{eq:ST2} \\
\fl	& = & \lim_{\Linf} Z^{-1} \int_{{\cal M}^{L}}\rmd 
\mu_{L}[\n] \rme^{-{\cal S}_{L}[\n]}\prod_{l=1}^{L}\left(1-\frac{\beta 
P_{H}(\n_{l})+\rmi\blambda\cdot {\bi 
P}(\n_{l})}{L}\right).\label{eq:ST3}
\end{eqnarray}
Here ${\cal M}^{L}$ comprises ordered sets of $L$ points
$\n_{l}\in{\cal M}$, with measure
$\rmd\mu_{L}[\n]=\prod_{l=1}^{L}\rmd\mu(\n_{l})$.  The 
gauge-invariant action ${\cal S}_{L}[\n]$ is defined by
\begin{equation}
	\exp(-{\cal S}_{L}[\n]) = \langle\n_{1}|\n_{2}\rangle 
	\langle\n_{2}|\cdots|\n_{L}\rangle 
	\langle\n_{L}|\n_{1}\rangle.
	\label{eq:expaction}
\end{equation}
We can consider these paths as a truncation of the function
space$\{\n:[0,\beta)\rightarrow {\cal M}\}$ to piecewise-constant
paths
\begin{equation}
	\n(\tau) = \n_{l}, \; (l-1)\beta/L \le \tau < 
	l\beta/L.
	\label{eq:ntaudisc}
\end{equation}
If we consider 
successive points as linked by a geodesic, with ${\bi n}_{L}$ linked 
to ${\bi n}_{1}$, these paths represent 
geodesic polygons.  The amplitude of this factor \eref{eq:expaction} decreases 
with increasing path length, and its (Berry)
phase is related to the enclosed area \cite{MS}.

\Eref{eq:ST3} can be re-exponentiated to 
give
\begin{equation}
\lim_{\Linf}\chi_{L}(\blambda) = \chi(\blambda)
	\label{eq:limchi}
\end{equation}
where
\begin{eqnarray}
	\chi_{L}(\blambda)&=&\left\langle \exp\left(-\rmi L^{-1}\sum_{l=1}^{L}\blambda\cdot{\bi 
P}(\n_{l})\right)\right\rangle_{L} \nonumber \\
&=&\left\langle \exp(-\rmi \blambda\cdot\overline{\bi 
P}[\n])\right\rangle_{L}.
	\label{eq:chiL}
\end{eqnarray}
Here the functional average is denoted 
\begin{equation}
\left\langle f\right\rangle_{L}\equiv Z^{-1} \int_{{\cal M}^{L}}\rmd 
	\mu_{L}[\n] f[\n]\exp\left(-{\cal S}_{L}[\n]-\beta 
	\bar{P_{H}}[\n]\right).
	\label{eq:aveL}
\end{equation}
and the \emph{path centroid} is
\begin{equation}
	\overline{P}[\n]=\frac{1}{L}\sum_{l=1}^{L}P(\n_{l}).
	\label{eq:centroid}
\end{equation}
The path centroid distribution (PCD) $W_{L}({\bi x})$, the approximant to the Wigner function 
obtained from polygonal paths with $L$ vertices, is the Fourier transform of 
$\chi_{L}$,
\begin{equation}
W_{L}({\bi x})=\left\langle \delta_{N}({\bi x}-\overline{\bi 
P}[\n])\right\rangle_{L}.
	\label{eq:WL}
\end{equation}
From \eref{eq:limchi} we obtain our main result,
\begin{equation}
\lim_{\Linf}W_{L}({\bi x})=W({\bi x}).
	\label{eq:limW}
\end{equation}
This limit exists in the sense of distributions; averages of polynomials 
in $\bi x$ with weight $W_{L}$ converge with $L$.  Since the Wigner function 
may be singular, as in equation 
\eref{eq:evenodd}, convergence of the PCD is not necessarily pointwise.

\subsection{Finite $L$ corrections}
\label{subsec.finite}
By cumulant expansion of the characteristic function \eref{eq:ST1} we
obtain finite-$L$ corrections to the Wigner function \eref{eq:WL} for 
the case of vanishing Hamiltonian.  Inclusion of the leading large-$L$
correction in \eref{eq:ST1} gives
\begin{equation}
	\chi(\blambda) = 
	Z^{-1}\Tr\left(1-\frac{1}{L}\rmi\lambda_{\mu}\hat{x}_{\mu}-
	\frac{1}{2L^{2}}\lambda_{\mu}\lambda_{\nu}\hat{x}_{\mu}\hat{x}_{\nu}
	+\Or(L^{-3})
	\right)^{L}.
	\label{eq:chiL3}
\end{equation}
(Summation over repeated indices is assumed.)  Replacing the factors by their diagonal representations, as in 
\sref{subsec:PCD}, and performing a cumulant expansion gives
\begin{equation}
\fl \chi(\blambda)=\left\langle
\exp\left(\frac{-\rmi\lambda_{\mu}}{L}
\sum_{l=1}^{L}P_{\mu}(\n_{l})-\frac{\lambda_{\mu}\lambda_{\nu}}{2L^{2}}\sum_{l=1}^{L}(P_{\mu\nu}(\n_{l})-
P_{\mu}(\n_{l})P_{\nu}(\n_{l}))+\Or(L^{-2})\right)\right\rangle_{L}.
	\label{eq:chiL2}
\end{equation}
Defining the  average
\begin{equation}
	C_{\mu\nu}=\left\langle \frac{1}{L}\sum_{l=1}^{L}\left(P_{\mu\nu}(\n_{l})-
P_{\mu}(\n_{l})P_{\nu}(\n_{l})\right)\right\rangle_{L},
	\label{eq:cumulant}
\end{equation}
and comparing \eref{eq:chiL2} with \eref{eq:chiL} gives
\begin{equation}
	\chi_{L}(\blambda) =
	\chi(\blambda)\exp\left(
	\frac{1}{2L}\lambda_{\mu}\lambda_{\nu} C_{\mu\nu}
	+ \Or(L^{-2})\right),
	\label{eq:chiLchi}
\end{equation}
provided that the quantity in angle brackets in \eref{eq:cumulant} is uncorrelated with the linear 
term.  In that case, if $C$ is 
negative-definite, we can compute the PCD
as the exact Wigner function convolved (to leading order) with a 
Gaussian of variance inversely proportional to $L$:
\begin{equation}
\fl	W_{L}({\bi x}) \approx (\det(-2\pi C/L))^{-1/2} \int \rmd^{N}{\bi y}\;W(\bi{ 
x-y})\exp\left(LC^{-1}_{\mu\nu}y_{\mu}y_{\nu}/2\right).
	\label{eq:Wsmear}
\end{equation}
It is worth noting that the distribution of auxiliary fields 
coupled to these operators $\hat\bi x$ is similarly the Wigner 
function convolved with a Gaussian of variance proportional to 
temperature \cite{S95}.

\section{Illustrative example: spin $s$}\label{sec:spin}

We conclude this section with the example of a spin $s$ 
particle with vanishing Hamiltonian, both to illustrate the theory 
and to recover and extend previous results \cite{S95,S99}.  

The path integration is now over spherical-polygon paths on the unit 
(Bloch) sphere.  Following  \eref{eq:overlap}, the action of such spherical-polygon paths is
\begin{equation}
	{\cal S}_{L} = -s\sum_{l=1}^{L}\ln \left( \frac{1+\n_{l-1}\cdot
	\n_{l}}{2} \right) -\rmi s\Omega,
	\label{eq:action}
\end{equation}
where $\Omega$ is the solid angle enclosed and $\n_{0}\equiv \n_{L}$.  
The imaginary part $s\Omega$ is the Berry phase.
The resulting PCD for a free spin $s$ is 
\begin{equation}
W_{L}({\bi S})=\left\langle \delta_{3}({\bi S}-(s+1)\nbar)\right\rangle_{L}.
	\label{eq:WLS}
\end{equation}
From the previous argument, this is the spin Wigner 
function \eref{eq:evenodd}, broadened by a Gaussian for large $L$.  To find the 
variance of this Gaussian, we insert into \eref{eq:cumulant} Lieb's
expressions \cite{Lieb} for the diagonal representatives of $\hat{S}_{z}$ and
$\hat{S}_{z}^{2}$, which are $P_{z}=(s+1)\cos\theta$ and
$P_{zz}=(s+1)(s+3/2)\cos^{2}\theta - (s+1)/2$ respectively:
\begin{equation}
	C_{\mu\nu}=-\frac{s+1}{3}\delta_{\mu\nu}.
	\label{eq:Scumulant}
\end{equation}
The PCD \eref{eq:Wsmear} is therefore the exact Wigner function \eref{eq:evenodd}
broadened by a Gaussian of 
variance $(s+1)/(3L)$:
\begin{equation}
\fl	W_{L}({\bi S}) \approx \frac{1}{2s+1}\left( \frac{3L}{2\pi (s+1)} \right)^{\frac{3}{2}}
	 \sum_{m=-s}^{s}\left(1-\frac{m}{S}\right) 
	 \exp\left(-\frac{3L(S-m)^{2}}{2(s+1)}\right) .
	\label{eq:WsmearS}
\end{equation}
\Eref{eq:WLS} 
implies the vanishing of the exact PCD for $S 
> s+1$ (i.e., $|\nbar| > 1$).  The approximation \eref{eq:WsmearS}
has Gaussian tails in this region, vanishing in the 
large-$L$ limit for any fixed $S>s$.

The expression \eref{eq:WLS} suggests a means of calculation of the 
PCD by accumulation of a histogram of values of $\nbar$.  While 
closed forms for the PCD are obtainable for small $L$ and an asymptotic 
form \eref{eq:WsmearS} for large $L$, it is nevertheless informative to compute the 
PCD by Monte Carlo integration.  For each Monte Carlo step, an ordered set of
$L$ independent unit vectors $\n_{l}$ is drawn from a uniform
distribution on the sphere and their vector average $\nbar$ taken. 
The real part of the weight $\exp(-{\cal S}_{L})$ is added to the
corresponding bin in the histogram.  As the time-reversed path appears
with equal probability, the imaginary part is discarded.  There is substantial cancellation of the 
real part of the weights (the sign problem), which hinders convergence.  

\begin{figure}
\begin{center}
\epsfig{file=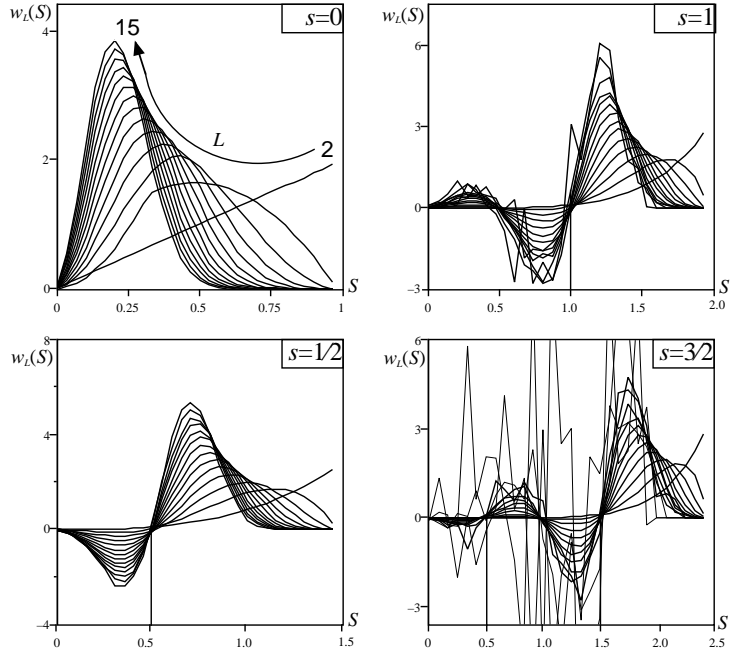,width=10cm}
\end{center}
\caption{Radial path centroid distribution \eref{eq:radial} for
spherical polygons with $L=2\cdots 15$ vertices, computed for spins
$s=0\cdots 3/2$ with $10^{7}$ Monte Carlo steps.  The vertical bars
show the expected positions of the $\delta^{\prime}$ functions as
$L\rightarrow\infty$.  To aid the eye, plots are shown as thin lines
where convergence is poor.
\label{fig:Wigner}}
\end{figure}

Since the PCD is spherically symmetric, we only require the radial distribution
\begin{equation}
	w_L(S) = 4 \pi S^{2} W_{L}({\bi S}).
	\label{eq:radial}
\end{equation}
Because the coordinate transformation is nonlinear, this distribution
is not the Wigner function of the operator $\sqrt{\hat{\bi
S}\cdot\hat{\bi S}}$; in particular, it is not a positive distribution.
\Fref{fig:Wigner} shows the numerically accumulated histograms for the radial 
distribution and their convergence with $L$ to the Wigner function.  The plots are shown for a
relatively small number $10^{7}$ of Monte Carlo steps to highlight
the influence of the sign problem on convergence.

\begin{figure}
\begin{center}
\epsfig{file=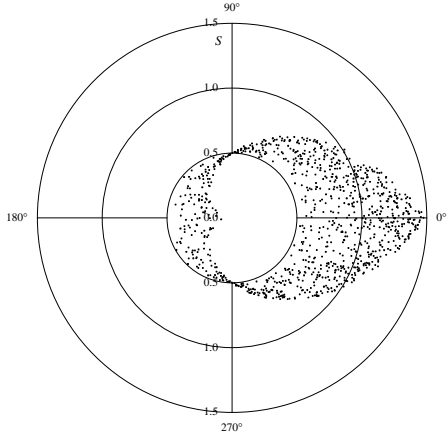,width=6cm} 
\end{center}
\caption{Polar plot of magnitude of path-centroid spin $S=3|\nbar|/2$
against Berry phase for spin $1/2$ for a random sample of $1000$ spherical
triangles.  The phase is less (greater) than $\pi/2$ in magnitude 
for $S>(<)1/2$.
\label{fig:phase}}
\end{figure}
Knowledge of the phase distribution, and of the correlation between
phase and path centroid, is of importance both to determine the
feasibility of convergence and to develop methods of accelerating
convergence \cite{Deutsch}.  The paths contributing to $|\nbar|
\approx 1$ enclose a small area, so that the phase distribution is
peaked near zero and the spin distribution is positive.  The action is
real for $L=1$ (the static approximation to the path integral), and
for $L=2$, where the path encloses no area.  Non-zero phases appear
from $L=3$, where the spherical polygons enclose non-zero area.  This
is the smallest value that retains information about spin
quantization, and distinguishes the spectra of ferromagnetic and
antiferromagnetic spin systems \cite{S00}.  In this case the weights are complex,
but there is still correlation between the time-averaged spin and the
phase, as \fref{fig:phase} shows.  For $L\gg 3$ the Berry phases
(modulo $2\pi$) are nearly uniformly distributed between 0 and $2\pi$
\cite{SS}, and weakly correlated with the path centroid, so that
convergence is poor.  Statistics are poor for the Wigner function for
large path averages $(|\nbar| \gg 1/\sqrt{L})$ due to uniform
sampling, and for small path averages $(|\nbar| \ll \frac{s}{s+1})$
due to destructive interference between paths of positive and negative
weights.  These ranges overlap for large $s$ and $L$, giving poor
statistics for the full interval.  The histograms would converge to
the Wigner function if the limit of infinitely many Monte Carlo steps
were taken \emph{before} the limit $L \rightarrow \infty$. 

\section{Discussion}
\label{sec:conc}
The main result of this investigation is proved in \sref{subsec:PCD}:
the Wigner function of a set of operators is obtained in terms of the
path centroid distribution (PCD) of the diagonal representatives of
the operators.  The Wigner function is obtained in the limit of
geodesic polygonal paths as the number of vertices $\Linf$. 
Quantization is apparent when $L\ge 3$ and becomes exact as $\Linf$. 
The distribution corresponding to commuting operators, measured at a
single time, is a non-negative effective Boltzmann distribution in
configuration space \cite{FH,GT,FK,CV}.  Interference between paths
occurs as the result of superposition of amplitudes, but the
probabilities are positive.  If the operators do not commute, the
resulting distribution need not be positive.  Should one interpret the
exponentiated action of a path not just as the weight by which
functionals of the path are to be averaged, but as a (complex)
quasiprobability that a spin take a particular closed path, one can
see the quasiprobabilities themselves interfering destructively.  The
question of whether incompatible observables each take given values is
impermissible in quantum mechanics; we should not expect the answer to
be drawn from a positive distribution.  However, the question of
whether the point $\bi{x}$ lies in a subset $\gamma \subset \Gamma$ is
well defined, provided that $\gamma$ is sufficiently large that the
coarse-grained Wigner function $\int_{\gamma}W({\bi x})\rmd{\bi x}$ is
positive.  Thus the sum of the weights of all paths with centroid in
$\gamma$ will be positive.

A heuristic argument helps one to understand the oscillations in sign of 
the Wigner function in terms of the Berry phases of paths.  Because long
paths are suppressed by an overlap factor \eref{eq:expaction}, the
dominant paths $\{{\bi P}({\bi n}_{l}), l=1\ldots L\}$ contributing to
the path centroid $\bar{\bi P}=\sum_{l}{\bi P}({\bi n}_{l})/L$ are the
shortest ones exploring a region of low energy.  The Wigner function
will be negative where these paths have a Berry phase near an odd
multiple of $\pi$.  Of course the entropy gain in extending the paths 
leads to a broad distribution of paths \cite{SS}, but the minimum may 
still be visible as a caustic.  There are two ways this may arise.  Firstly, if
the coherent state manifold is curved the path centroid does not in
general lie in the manifold; this implies a lower bound on the length of
the path.  Such is the case for SU(2) coherent states discussed in
\sref{sec:spin}.  The shortest paths on the unit sphere with 
centroid $|\nbar|=\cos\theta$ are small circles enclosing solid 
angle $2\pi(1-\cos\theta)$; a spin $s$ performing such a conical path 
will acquire a Berry phase $2\pi s(1-\cos\theta)$.  The PCD should 
therefore be positive for $\cos\theta=1$ and cross zero whenever the 
phase is an odd multiple of $\pi/2$.  Such an argument predicts zeros 
with positive gradient at 
\begin{equation}
	\cos\theta=\frac{4s-1}{4s}, 
	\frac{4s-5}{4s} \ldots \frac{3}{4s}\left[\frac{1}{4s}\right],
	\label{eq:heuristiczeros}
\end{equation}
to be compared with the exact result \eref{eq:evenodd}
\begin{equation}
	\cos\theta=\frac{s}{s+1}, 
	\frac{s-1}{s+1} \ldots \frac{1}{s+1}\left[ \frac{1}{2(s+1)}\right],
	\label{eq:exactzeros}
\end{equation}
for integer [half-integer] spins respectively.  This (admittedly
crude) argument gives the correct spacing of the zeros to $\Or(1/s)$. 
It is also possible for the dominant paths to have non-zero phase in a
flat coherent state manifold if Hamiltonian has a local maximum near
the position of the path centroid.  This is left for a further
investigation.

These results provide insight into the origin of the Monte Carlo sign problem.  If
the variables sampled in a quantum Monte Carlo simulation correspond 
to non-commuting operators, the joint
distribution of their time average will be a Wigner function,
indicating the presence of a non-positive integrand.  Such an example
might be the total spin of an interacting system or, more generally, a
vector order parameter generating a non-Abelian symmetry group.  Much more frequently,
the Hubbard-Stratonovich transformation replaces the interaction by an
auxiliary field, with Monte Carlo integration over time-dependent
configurations of this field.  The distribution of the time-averaged auxiliary field is
a Gaussian convolution of the distribution of the operators to which
it is coupled, with variance proportional to temperature \cite{S95}. 
Thus the same considerations apply to the auxiliary-field Monte Carlo
method at low temperatures; however, as the path space differs from 
the geodesic polygons discussed here, the results will be 
quantitatively different.


\section*{References}
\begin{thebibliography}{99}
	\bibitem{Feynman} Feynman R P 1987 in \textit{Quantum Implications} 
	edited by Hiley B J and Peat F D (London: Routledge) p~235
	\bibitem{Strat} Stratonovich R L 1956 \textit{Zh.  Eksp.  Teor. 
	Fiz.} {\bf 31} 1012 [1957 \textit{Sov.  Phys.  JETP} {\bf 4} 891]
	\bibitem{HOSW} Hillery M, O'Connell R F, Scully M O and Wigner E P
	1984 \textit{Phys.  Rep.} \textbf{106} 121
	\bibitem{Mueckenheim}  M\"{u}ckenheim W \etal 1986 \textit{Phys. Rep.} 
	\textbf{133} 337
	\bibitem{Bell} Bell J S 1964 \textit{Physics} \textbf{1} 195
	\bibitem{BSS} Blankenbecler R, Scalapino D J and Sugar R L 1981 \textit{Phys. 
		Rev. D} {\bf 24} 2278
	\bibitem{review} Samson J H 1995 \textit{Int. J. Mod. Phys. C} \textbf{6} 
	427 
	\bibitem{Takano} Takano H 1987 in \textit{Quantum Monte Carlo Methods in 
	Equilibrium and Non-equilibrium Systems} edited by 
	Suzuki M (Berlin: Springer) p~144
	\bibitem{VS2} Vieira V R and Sacramento P D 1994 \textit{Physica A} 
	\textbf{207} 584 
	\bibitem{Fahy} Fahy S and Hamann D R 1991 \textit{Phys. Rev. B} {\bf 43} 765 
	\bibitem{S95}  Samson J H 1995 \textit{Phys. Rev. B} \textbf{51} 223 
	\bibitem{FH} Feynman R P and Hibbs A R 1965 \textit{Quantum Mechanics 
	and Path Integrals} (New York: McGraw-Hill) p~280
	\bibitem{GT} Giachetti R and Tognetti V 1985 \textit{Phys. Rev. Lett.}
	\textbf{55} 912 
	\bibitem{FK} Feynman R P and Kleinert H 1986 \textit{Phys. Rev. A} \textbf{34} 
	5080 
	\bibitem{CV}  Cao J and Voth G A 1994 \textit{J. Chem. Phys.} \textbf{100} 
	5093 
	\nonum  Cao J and Voth G A 1994 \textit{J. Chem. Phys.} \textbf{100} 
	5106 
	\bibitem{CGTVV}  Cuccoli A \etal 1995 \textit{J. Phys: Condens. Matter} 
	\textbf{7} 7891 
	\bibitem{FS} Fradkin E and Stone M 1988 \textit{Phys. Rev. B} {\bf 38} 7215 
	\bibitem{Klauder} Klauder J R 1982 \textit{J. Math. Phys.} \textbf{23} 
	1797 
	\bibitem{DK} Klauder J R and Daubechies I 1984 \textit{Phys. Rev. Lett.} 
	\textbf{52} 1161 
	\nonum Daubechies I and Klauder J R 1985 \textit{J. Math. Phys.} 
 	\textbf{26} 2239 
	\bibitem{Prange} Prange R E 1981 in \textit{Electron Correlation and 
	Magnetism in Narrow-band Systems} edited by 
	Moriya~T (Berlin: Springer) p~69
 	\bibitem{S99} Samson J H 1999 in \textit{Proceedings of the Sixth International
	Conference on Path Integrals from peV 
	to TeV, Firenze} edited by Casalbuoni R \etal (Singapore: World 
	Scientific) p~550, quant-ph/9905089
	\bibitem{SW} Scully M O and W\'{o}dkiewicz K 1994 \textit{Found. Phys.} 
	\textbf{24} 85 
	\bibitem{Agarwal} Agarwal G S, Home D and Schleich W 1992 \textit{Phys. Lett.
		A} {\bf 170} 359 
	\nonum Agarwal G S 1993 \textit{Phys. Rev. A} \textbf{47} 4608 
	\bibitem{Chandler} Chandler C \etal 1992 \textit{Found. Phys.}
	\textbf{22} 867 
	\bibitem{KS} Klauder J R and Skagerstam B S 1994 \textit{Coherent 
	States: Applications in Physics and Mathematical Physics} (Singapore: World 
	Scientific)
	\bibitem{Perelomov} Perelomov A 1986 \textit{Generalized Coherent 
	States and their Applications} (Berlin: Springer) p~59
	\bibitem{ZFG} Zhang W M, Feng D H and Gilmore R 1990 \textit{Rev. Mod. 
		Phys.} {\bf 62} 867 
	\bibitem{BM} Brif C and Mann A 1999 \textit{Phys. Rev. A} \textbf{59} 971 
	\bibitem{MACS} Mukunda N \etal, quant-ph/0002070
 	\bibitem{VS} Vieira V R and Sacramento P D 1994 \textit{J. Phys. A: Math. Gen.}
 	\textbf{27} L783 
 	\bibitem{Gilmore} Gilmore R 1976 \textit{J. Phys. A} \textbf{9} L65 
	\bibitem{Radcliffe} Radcliffe J M 1971 \textit{J. Phys. A: Math. Gen.} \textbf{4} 313 
	\bibitem{Lieb} Lieb E H 1973 \textit{Commun. Math. Phys.} \textbf{31} 327 
	\textbf{132} 277 
	\bibitem{MS} Mukunda N and Simon R 1993 \textit{Ann. Phys., NY} \textbf{228} 
	205 
	\bibitem{Deutsch}  Deutsch J M 1994 \textit{Phys. Rev. E} \textbf{50} R2411 
	\bibitem{S00} Samson J H 2000 \textit{J. Phys. A} \textbf{33} 3111
	\bibitem{SS} Sinha S and Samuel J 1994 \textit{ Phys. Rev. B} \textbf{50} 13871 
\end{thebibliography}
\end{document}